\documentclass{epl}

\usepackage{graphicx}
\usepackage{epstopdf}
\usepackage{amsmath}

\vspace*{-1cm}

\title{Membrane adhesion {\em via} competing receptor/ligand bonds} 
\shorttitle{\small Competing receptor/ligand bonds}
\author{M.\ Asfaw \and B.\ R\'{o}\.{z}ycki \and R.\ Lipowsky \and T.\ R.\ Weikl}
\institute{Max Planck Institute of Colloids and Interfaces,  14424 Potsdam, Germany}


\begin{document}

\maketitle

\vspace*{-0.5cm}

\begin{abstract}
The adhesion of biological membranes is controlled by various types of receptor and ligand molecules. In this letter, we present a statistical-mechanical model for membranes that interact via receptor/ligand bonds of two different lengths. We show that the equilibrium phase behavior of the membranes is governed by an effective double-well potential. The depths of the two potential wells depend on the concentrations and binding energies of the receptors and ligands. The membranes are unbound for small, and bound for larger potential depths. In the bound state, the length mismatch of the receptor/ligand bonds can lead to lateral phase separation. We derive explicit scaling laws for the critical points of unbinding and phase separation, and determine the prefactors by comparison with Monte Carlo results. 
\end{abstract}

\vspace*{-0.5cm}

\section{Introduction}

Biological membranes consist of a lipid bilayer with embedded or adsorbed proteins \cite{Alberts02}. The adhesion of the membranes is typically mediated by various types of receptor and ligand proteins. Each type of receptor specifically binds to a complementary type of ligand in the apposing membrane. The adhesion of immune cell membranes is often mediated by receptor/ligand bonds of different length. The receptor/ligand complexes that mediate the adhesion of T cells, for example, have characteristic lengths of 15 or 40 nm \cite{Dustin00}. During T cell adhesion, a lateral phase separation into domains that are either rich in short or long receptor/ligand bonds has been observed in the cell contact zone \cite{Monks98,Grakoui99}. The domain formation is assumed to be driven by the length mismatch of the receptor/ligand bonds \cite{Qi01,Weikl02,Burroughs02,Raychaudhuri03,Chen03,Coombs04,Weikl04}. 

We consider here a statistical-mechanical model of two membranes interacting via long and short receptor/ligand bonds. In our model, the membranes are discretized into small patches that can contain single receptor or ligand molecules. The conformations of the membranes are described by the local separation of apposing membrane patches, and by the distribution of receptors and ligands in the membranes. We show that a summation over the receptor and ligand degrees of freedom in the partition function leads to an effective double-well potential. The potential well at small membrane separations reflects the interactions of the short receptor/ligand bonds, and the potential well at larger separations the interactions of the long receptor/ligand bonds. The depths of the wells depend on the concentrations and binding energies of the receptors and ligands.

We focus on the equilibrium phase behavior of the membranes, which exhibit two characteristic phase transitions. The first transition is the unbinding transition of the membranes, which is driven by an entropic membrane repulsion arising from thermal shape fluctuations. The second transition is lateral phase separation within the membranes,  driven by the length mismatch of the receptor/ligand bonds. The length mismatch leads to a membrane-mediated repulsion between the two different receptor/ligand bonds, because the membranes have to be bent to compensate this mismatch, which costs elastic energy. This repulsion leads to a lateral phase separation for sufficiently large concentrations of the receptor/ligand bonds and, thus, sufficiently deep wells of the effective potential.

For both types of phase transitions, we derive characteristic scaling laws. We confirm these scaling laws using Monte Carlo simulations, and obtain the numerical prefactors by a comparison with the simulation results. From the scaling arguments and simulations, we thus obtain relatively simple relations that characterize the phase behavior of the membranes. 

\section{Model and effective potential}
 
\begin{figure}[t]
\begin{center}
\resizebox{0.55\columnwidth}{!}{\includegraphics{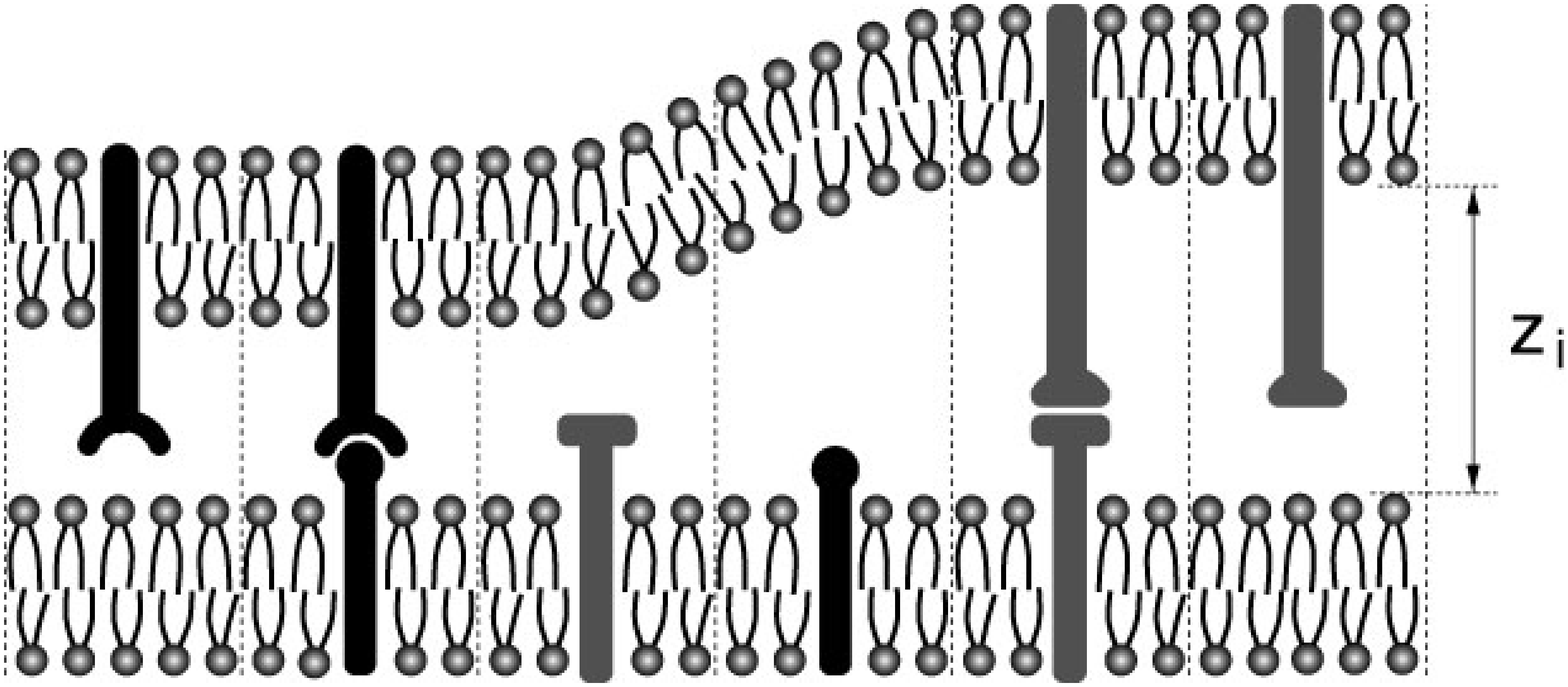}}
\hspace{0.0cm}
\resizebox{0.43\columnwidth}{!}{\includegraphics{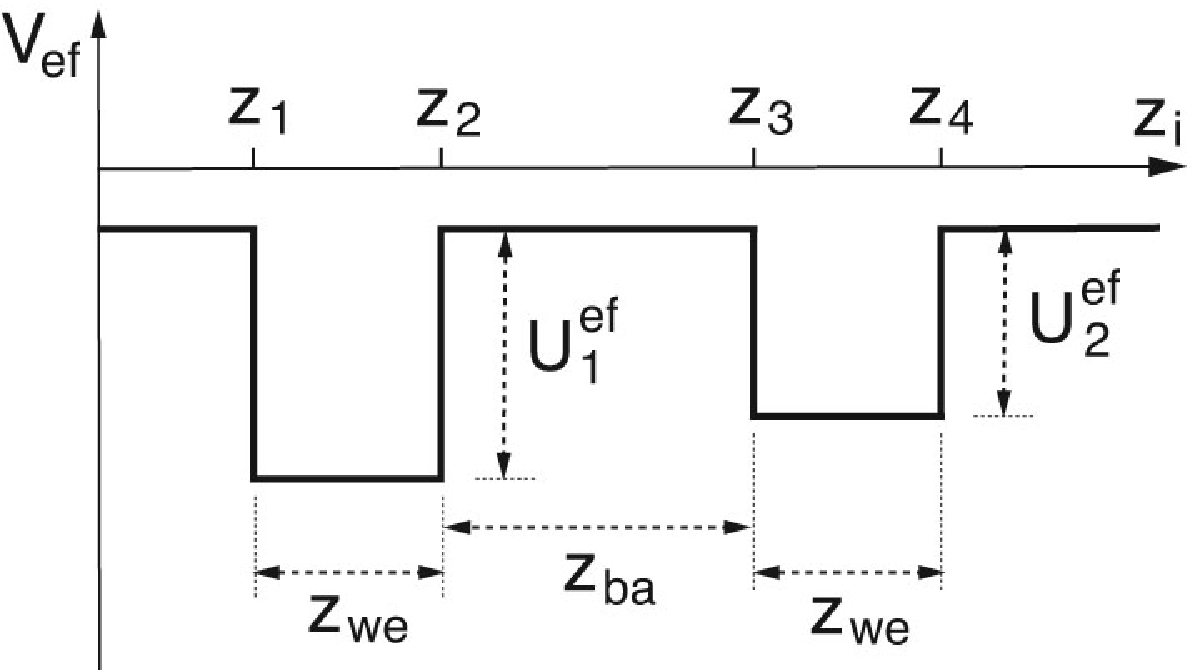}}
\caption{(Left) A membrane containing long and short receptor molecules (top) adhering to a membrane with complementary ligands. In our model, the membranes are discretized into small patches (indicated by dashed lines). A membrane patch can contain a single receptor or ligand molecule. The membrane conformations are described by the local separation $z_i$ of each pair $i$ of apposing membrane patches. -- (Right) Summing over all possible distributions of receptor and ligand molecules in the partition function leads to an effective double-well potential $V_\text{ef}$ for the membranes. The potential well at short separations $z_1<z_i<z_2$ reflects the interactions of the short receptor/ligand bonds, the well at larger separations $z_3<z_i<z_4$ reflects the interactions of the long receptor/ligand bonds.}
\label{figure_cartoon}
\label{figure_potential}
\end{center}
\end{figure}
 
We consider two interacting membranes. One of the membranes contains long and short receptors, the other membrane contains complementary ligands (see Fig.~\ref{figure_cartoon}). We use a lattice gas model and discretize the membranes into quadratic patches of size $a\times a$ \cite{Lipowsky96}. Simulations with molecular membrane models indicate that the smallest bending deformation corresponds to a linear patch size $a\simeq 5$~nm \cite{Goetz99}. Each membrane patch can contain a single receptor or ligand molecule. We describe the distribution of receptors by a composition field $n_i$ with values 0, 1, or 2. The value $n_i=1$ indicates that a short receptor is present in patch $i$, $n_i=2$ indicates a long receptor, and $n_i=0$ indicates that no receptor is present. The distribution of ligands in the apposing membrane is described by the composition field $m_i$, with $m_i=1$, 2, or 0 indicating ligands of short receptors, ligands of long receptors, or no ligands in patch $i$ of this membrane. The membrane conformations can be described by the local separation $l_i$ of the apposing membrane patches $i$. 

The configurational energy of the membranes is the sum of elastic and interaction energies:
\begin{equation}
{\cal H}\{z,n,m\} = {\cal H}_\text{el}\{z\} + {\cal H}_\text{int}\{z,n,m\}\,, \text{~with~} {\cal H}_\text{el}\{z\}= \sum_i {\textstyle \frac{1}{2}} \left(\Delta_d z_i\right)^2 \label{Hamiltonian}
\end{equation}
To simplify the notation, we use the rescaled separation field $z_i = (l_i/a)\sqrt{\kappa/k_BT}$, where $\kappa = \kappa_1\kappa_2/(\kappa_1+\kappa_2)$ is the effective bending rigidity of the membranes with rigidities $\kappa_1$ and $\kappa_2$ \cite{Lipowsky88}. The discretized Laplacian
$\Delta_{d}z_i =z_{i1}+z_{i2}+z_{i3}+z_{i4}-4 z_i$ is proportional to the mean curvature of the separation field.
Here, $z_{i1}$ to $z_{i4}$ are the membrane separations at the four nearest-neighbor patches of membrane patch $i$. 
The interaction energy has the form
\begin{equation}
{\cal H}_\text{int}= \sum_i \big(\delta_{n_i,1}\delta_{m_i,1} V_1(z_i) + \delta_{n_i,2}\delta_{m_i,2} V_2(z_i)
-\; \delta_{n_i,1} \mu_{R1} -  \delta_{n_i,2} \mu_{R2} - \delta_{m_i,1} \mu_{L1} -  \delta_{m_i,2} \mu_{L2}
\big)
\end{equation}
where $V_1(z_i)$ is the interaction potential of a short receptor with its ligand, and $V_2(z_i)$ is the interaction potential of long receptors. The Kronecker symbol $\delta_{i,j}$ is equal to 1 for $i=j$, and equal to 0 for $i\neq j$. The term $\delta_{n_i,1}\delta_{m_i,1}$, for example, is only 1 for $n_i=1$ and $m_i=1$, i.e.~when a short receptor and a complementary ligand are present in the two apposing membrane patches labeled by $i$. The chemical potentials of the short and long receptors are denoted by $\mu_{R1}$ and  $\mu_{R2}$, and the chemical potentials of the ligands by $\mu_{L1}$ and $\mu_{L2}$. To simplify the notation, all energetic quantities, such as the configurational energy, interaction potentials and chemical potentials, are given in units of the thermal energy $k_B T$.

The equilibrium behavior of the membranes is governed by the partition function $\cal Z$. An exact summation over the degrees of freedom of the composition fields $n_i$ and $m_i$ in the partition function leads to
\begin{equation}
{\cal Z} \equiv \left[\prod_i \int_0^\infty \!\!\text{d}z_i\right] \left[\prod_i \sum_{n_i=0}^2\right] \left[\prod_i\sum_{m_i=0}^2\right] e^{-{\cal H}\{z,n,m\}} =\left[\prod_i \int_0^\infty \!\! \text{d}z_i\right] e^{-\left({\cal H}_\text{el}\{z\} + \sum_i V_\text{ef}(z_i)\right)} \label{Z}
\end{equation}
with the effective potential
\begin{equation}
\begin{split}
V_\text{ef}(z_i) =& - \ln\Big[ \zeta_0 + e^{\mu_{R1}+\mu_\text{L1}}\big(e^{-V_1(z_i)}-1\big)
+ e^{\mu_{R2}+\mu_{L2}}\big(e^{-V_2(z_i)}-1\big) \Big]
\end{split}
\label{Vef}
\end{equation}
and $\zeta_0\equiv \left(1 + e^{\mu_{R1}} + e^{\mu_\text{R2}}\right)\left(1 + e^{\mu_\text{L1}} + e^{\mu_\text{L2}}\right)$. Here, we characterize the molecular interactions by the square-well potentials 
$V_1(z_i) = -U_1$ for $z_1< z_i < z_2$ and  0 otherwise, and $V_2(z_i) = -U_2$ for $z_3< z_i < z_4$ and 0 otherwise. For $z_2 < z_3$, the effective potential $V_\text{ef}(z_i)$ is the double-well potential  shown in Fig.~\ref{figure_potential} with the effective depths 
\begin{equation}
U_1^\text{ef} =  \ln\Big[1 + e^{\mu_{R1}+\mu_\text{L1}}\big(e^{U_1}-1\big)/\zeta_0\Big] \mbox{~and~}  U_2^\text{ef} =  \ln\Big[1 + e^{\mu_{R2}+\mu_\text{L2}}\big(e^{U_2}-1\big)/\zeta_0\Big] \label{Uefs}
\end{equation}
of the two potentials wells with respect to $V_\text{ef}(z_i=\infty)=-\ln \zeta_0$.

\begin{figure}[t]
\begin{center}
\resizebox{\columnwidth}{!}{\includegraphics{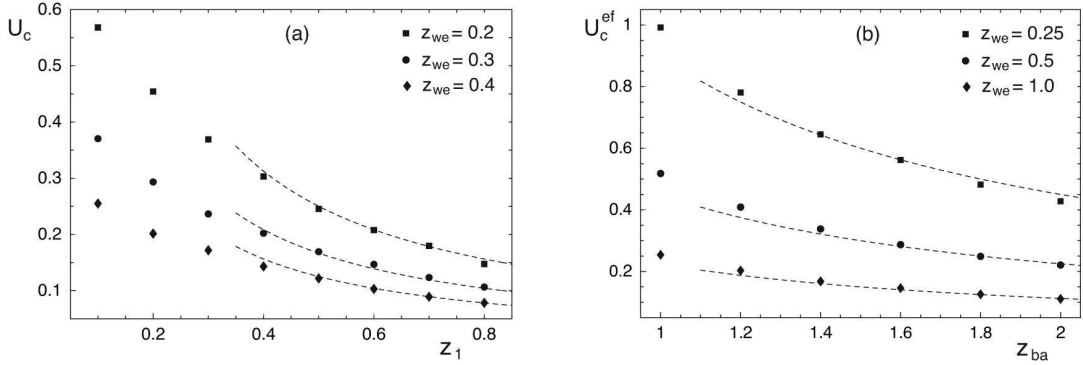}}
\caption{Monte Carlo data for membranes: (a) Critical potential depth $U_c$ of unbinding from the single-well potential (\ref{singlewell}); (b) Critical depth $U_c^\text{ef}$ for lateral phase separation in the symmetric double-well potential. The dashed lines represent the scaling laws (\ref{unbinding}) and (\ref{Uc_ef}) with numerical prefactors $b=0.025\pm 0.002$ and $c = 0.225 \pm 0.02$. The scaling laws are valid for large values of $z_1$, $z_\text{we}$ and $z_\text{ba}$. The given value of $b$ has been obtained from a fit to the Monte Carlo data points for $z_1\ge 0.5$ in (a), and the value of $c$ is obtained  from a fit to the data points for $z_\text{we} \ge 0.5$ and $z_\text{ba} \ge 1.4$ in (b). 
}
\label{figure_MCdata}
\end{center} 
\end{figure}

\section{Unbinding from single-well potential}

We first consider the single-well potential
\begin{equation}
V(z) = - U \text{~for~} z_1 < z < z_2 \; \text{~and~} V(z) = 0 \text{~otherwise} \label{singlewell}
\end{equation}
A potential of this form is obtained by setting $U_1^\text{ef}$ or $U_2^\text{ef}$ equal to 0 for the double-well potential shown in Fig.~\ref{figure_potential}. The critical scaling behavior of membranes interacting via such a potential is similar to the scaling behavior of strings. Since strings can be studied with analytical methods, we first determine the critical potential depth $U_c$ of strings as a function of $z_1$ and $z_2$  and subsequently compare with Monte Carlo results for membranes. 

Strings are lines governed by tension. We consider here two interacting strings in two dimensions. The conformations of the strings can be described by the local separation $l(x)$ perpendicular to a reference line. The strings are, on average, parallel to this line. To simplify the notation, we use again a rescaled separation field $z(x)=l(x)\sqrt{\sigma/k_B T}$ where $\sigma$ is the effective tension of the two strings. The effective Hamiltonian has the form \cite{Lipowsky88b}
\begin{equation}
{\cal H}\left\{ z \right\} =  \int \text{d}x\left[ \frac{1}{2} \left( \frac{\text{d}z}{\text{d}x}\right)^2 
+V(z) \right] 
\end{equation}
where $V(z)$ is the interaction potential of the strings. In the continuum limit, the separation of two interacting strings is equivalent to the spatial coordinate of a quantum-mechanical particle in one dimension \cite{Lipowsky88b}. The free energy of the strings then corresponds to the ground-state eigenvalue $E_0$ of the Schr\"odinger-type equation 
\begin{equation}
-{\partial^2 \psi_{k}\over \partial z^2}+V(z) \psi_{k}(z) = E_{k} \psi_{k}(z)   \label{S_equation}
\end{equation}
In the case of the square-well potential (\ref{singlewell}), the eigenfunction $\psi_0$ has the form 
\begin{equation}
\begin{split}
\psi_{0}(z) & =  A_{1} (\exp (k z) - \exp (-k z))  \text{~for~}  0 < z < z_1 \\
         & = A_{2} \cos (\alpha z) + A_{3} \sin (\alpha z)  \text{~for~}  z_1 < z < z_2 \\
         & = A_{4} \exp (-k z)  \text{~for~} z>z_{2}  \label{eigenfunction}
\end{split}
\end{equation}
with $ \alpha^2 = E_0 + U$ and $k^{2} = - E_{0}$. Since the strings cannot penetrate each other, we have $\psi_{0}(z)=0$ for $z<0$. The two strings are bound for $E_0<0$, and unbind at $E_0=0$. The eigenfunction $\psi_{0}(z)$ and its derivative $\text{d}\psi_{0}(z)/\text{d}z$ have to be continuous at $z=z_1$ and $z=z_2$. These four continuity conditions lead to a transcendental equation. At the unbinding point, i.e.~at $E_0=0$,  the transcendental equation is $\cos \left[ (z_2 - z_1) \sqrt{U_c} \right] = z_{1} \sqrt{U_c} \sin \left[ (z_{2} - z_{1}) \sqrt{U_c}\right]$. 
For $z_1=0$, this equation simplifies to $\cos(z_2 \sqrt{U_c}) =0$, which implies $U_c = \pi^2/( 4 z_2^2)$.  For $z_1 \neq 0$, the equation can be rearranged to $z_1 \sqrt{U_c} \tan \left[ (z_2-z_1) \sqrt{ U_c} \right] =1$, which leads to 
\begin{equation}
U_c = \frac{b}{z_1(z_2-z_1)} \text{~~~for~} (z_2-z_1) \sqrt{ U_c} \ll 1\label{unbinding}
\end{equation}
with prefactor $b=1$.

Functional renormalization arguments indicate that membranes have the same critical scaling behavior as strings \cite{Lipowsky88}. It is therefore reasonable to assume that the critical potential depth $U_c$ of two membranes interacting via the single-well potential (\ref{singlewell}) is governed by the same scaling law (\ref{unbinding}), but with a prefactor $b$ that is different from the factor $b=1$ for strings. To test this proposition, we determine the critical potential depth for membranes with Monte Carlo simulations. In the simulations, we use the discretized Hamiltonian ${\cal H}\{z\} = {\cal H}_\text{el}\{z\} + \sum V(z_i)$ with the elastic energy (\ref{Hamiltonian}) and the single-well potential (\ref{singlewell}), and attempt local moves in which the separation $z_i$ of patch $i$ is shifted to a new value $z_i + \zeta$ where $\zeta$ is a random number between $-1$ and 1. Following the standard Metropolis criterion \cite{Binder92}, a local move is always accepted if the change $\Delta {\cal H}$ in conformational energy is negative, and accepted with the probability $\exp(-\Delta  {\cal H})$ for $\Delta {\cal H}>0$. We perform simulations with up to $10^7$ attempted local moves per site $i$ and membrane sizes of $N=120 \times 120$ patches. At the unbinding point, the correlation length and the autocorrelation time of the membranes diverge towards infinity. To determine the critical potential depth $U_c$, we perform simulations for larger potential depths $U>U_c$ at which the lateral correlation length is significantly smaller than the membrane size. Thermodynamic averages $\langle \bar{z}\rangle$ and $\langle P_b\rangle$ for the mean separation $\bar{z}=\frac{1}{N}\sum_{i=1}^N z_i$ and the fraction $P_b$ of membrane patches bound in the potential well then do not depend on the finite system size. The critical potential depth $U_c$ is obtained from extrapolating $\langle \bar{z}\rangle$ and $\langle P_b\rangle$ as functions of $U$ to the critical values $1/\langle \bar{z}\rangle=0$ and $\langle P_b\rangle=0$. Monte Carlo results for $U_c$ at various values of the potential lengths $z_1$ and $z_2$ are shown in Fig.~\ref{figure_MCdata}(a). Since the membranes are discretized, we only expect agreement with eq.~(\ref{unbinding}) in the continuum limit of large $z_1$ and $z_2$. From fitting eq.~(\ref{unbinding}) to the Monte Carlo results for large $z_1$ and $z_2$, we obtain $b= 0.025\pm 0.002$.

\section{Lateral phase separation in a symmetric double-well potential}

We now consider the case where the two wells of the effective potential shown in Fig.~\ref{figure_potential} are located at large membrane separations and neglect the hard-wall repulsion at $z=0$. By comparing with the results of the previous section, we will later show that this case is realistic for the typical dimensions of cell receptor/ligand bonds. We assume here that the two potential wells have the same width $z_\text{we}$. For symmetry reasons, the lateral phase separation then occurs at the same depth of the two potential wells. Therefore, we focus here on the symmetric double-well potential with $U^\text{ef}=U^\text{ef}_1=U^\text{ef}_2$. Our goal is to determine the critical potential depth $U_c^\text{ef}$ for lateral phase separation as a function of the width $z_\text{we}$ and separation $z_\text{ba}$ of the potential wells. 

To derive a scaling relation for $U_c^\text{ef}$, we first show that the free energy of membranes bound in a single-well potential of depth $U$ and width $z_\text{we}$ scales as
\begin{equation}
{\cal F} \sim (U z_\text{we})^2 \label{F_singlewell}
\end{equation}
for small values of $U$ and negligible hard-wall repulsion. This scaling law can be obtained again from the analogy with strings. If the hard wall is negliglible, the ground-state eigenfunction $\psi_0$ is symmetric with respect to the center $z_m = (z_2 + z_1)/2$ of the single well. The eigenfunction given in eq.~(\ref{eigenfunction}) then simplifies to $\psi_0(z) = B_1 \exp(-k|z-z_m|)$ for $z<z_1$ or $z>z_2$, and  $\psi_0(z) = B_2 \cos(\alpha (z-z_m))$ for $z_1 < z < z_2$. From the continuity conditions at $z_1$ and $z_2$, we obtain the transcendental equation
$\sqrt{E_0+U}\tan\left({\textstyle\frac{1}{2}}\sqrt{E_0+U}z_{we}\right)=\sqrt{-E_0}$. For small values $U$, the free energy $E_0$ of the strings is also small. We thus can expand the tangent in the transcendental equation and obtain the explicit solution $E_0=(-2- U z_\text{we}^2+2\sqrt{1+U z_\text{we}^2})/z_\text{we}^{2} \simeq - (U z_\text{we})^2/4$ to leading order in $U$, and, thus, the scaling law for the free energy ${\cal F}=E_0$ as in (\ref{F_singlewell}).

We now come back to the symmetric double-well potential with well depth $U^\text{ef}$. According to (\ref{F_singlewell}), the free energy of the state where the membranes are bound in one of the potential wells scales as ${\cal F} \sim (U^\text{ef}z_\text{we})^2$. More precisely, ${\cal F}$ is the free energy difference with respect to the unbound state. This free energy difference includes the entropy loss of the bound membranes, and is therefore different from the binding energy $U^\text{ef}$ for the wells. The absolute value, $|{\cal F}|$, corresponds to a free energy barrier that the membranes have to cross from one well to the other well. A central, previous result is that the membranes exhibit a lateral phase separation if this free energy barrier exceeds a critical barrier height ${\cal F}_c \sim 1/z_\text{ba}^2$ \cite{Lipowsky94}. From this scaling relation for the critical barrier height and eq.~(\ref{F_singlewell}), we obtain the scaling law
\begin{equation}
U_c^\text{ef} = \frac{c}{z_\text{we}z_\text{ba}} 
\label{Uc_ef}
\end{equation}
for the critical depth of the double-well potential.

\begin{figure}[t]
\begin{center}
\resizebox{0.95\columnwidth}{!}{\includegraphics{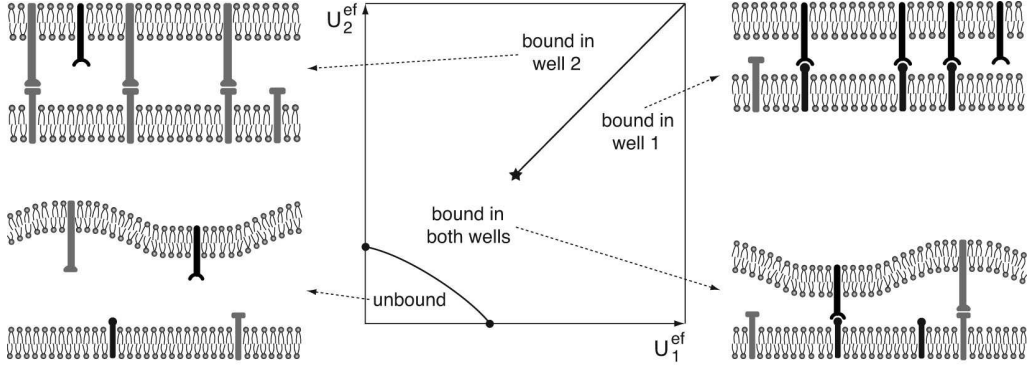}}
\caption{Phase diagram of membranes adhering via long and short receptor/ligand bonds. The membranes are unbound for small well depths $U_1^\text{ef}$ and $U_2^\text{ef}$ of the effective interaction potential shown in Fig.~\ref{figure_potential}, i.e.~for small concentrations or binding energies of receptors and ligands (see
eq.~(\ref{Uefs})). At large values of $U_1^\text{ef}$ and $U_2^\text{ef}$, the membranes are either bound in well 1 or well 2, i.e.~they are either bound by the short or by the long receptor/ligand bonds. At intermediate well depths $U_1^\text{ef}$ and $U_2^\text{ef}$, the membranes are bound in both potential wells. The critical point for the lateral phase separation (star) follows from eq.~(\ref{Uc_ef}). For typical dimensions of cell receptors and ligands, the critical well depth $U_c^\text{ef}$ for lateral phase separation is significantly larger than the critical depths of unbinding (see text). The critical unbinding points for $U_1^\text{ef}=0$ or $U_2^\text{ef}=0$ (dots) follow from eq.~(\ref{unbinding}).   }
\label{figure_phasediagram}
\end{center} 
\end{figure}

To verify eq.~(\ref{Uc_ef}) and to determine the numerical prefactor $c$, we compare again with Monte Carlo results. The critical potential depth can be determined from the moments $C_2 = \langle\bar{z}^2\rangle/\langle |\bar{z}|\rangle^2$, and
$C_4 = \langle\bar{z}^4\rangle/\langle \bar{z}^2\rangle^2$. Here, $\bar{z} = \frac{1}{N}\sum_{i=1}^N z_i$  is the spatially averaged separation, and  $\langle\cdots\rangle$ denotes an average over all membrane configurations. The values of these moments depend on the correlation length $\xi$ and the linear size $L$ of a given membrane segment.  At the critical point however, the correlation length $\xi$ diverges, and the values of the moments become independent of $L$ \cite{Binder92}. The critical depth $U^\text{ef}_c$ of the symmetric double-well potential can be estimated from the common intersection points of the functions $C_2(U^\text{ef})$ and $C_4(U^\text{ef})$ at different values of $L$ \cite{Binder92, Ammann96,Weikl02b}. We have considered quadratic membrane segments with a linear size of $L = 10$, 20, and 40 patches and periodic boundary conditions. From the comparison with the simulation results shown in Fig.~\ref{figure_MCdata}(b), we obtain $c=0.225\pm 0.02$.
 
In the present study, we ignored possible effects of membrane tension, $\sigma$, which dominates over the bending energy on length scales larger than the crossover length $\sqrt{\kappa/\sigma}$, but is negligible on smaller length scales. For lipid bilayers with a bending rigidity $\kappa$ of about 20  $k_B T$ and tensions of a few $\mu$J/m \cite{Simson98}, the crossover length $\sqrt{\kappa/\sigma}$ attains values of several hundred nanometers, which is significantly larger than the membrane deformations between domains of short and long cell receptor/ligand bonds \cite{Weikl04}. Therefore, tensions up to a few $\mu$J/m should only weakly affect the critical potential depth (\ref{Uc_ef}) for phase separation. However, tension prevents unbinding from the short-range potentials considered here \cite{Lipowsky96}. In addition, active switching processes of stickers can have a strong impact on the unbinding behavior \cite{Rozycki06}.
 
\section{Phase diagram and Conclusions} 
 
In the case of T cells, the long and short receptor/ligand bonds have linear extensions of $l=40$ and 15 nm, respectively \cite{Dustin00}. Reasonable values for the width of the potential wells range from $l_\text{we} = 1$ to 5 nm. The rescaled width $z_\text{we}$ of the two potential wells is then significantly smaller than rescaled separation $z_1$ of the first well from the hard wall and the separation $z_\text{ba}$ of the two wells, and the ratio of $z_1$ and $z_\text{ba}$ is around 0.6. For these values, the critical potential depth $U_c^\text{ef}$ for lateral phase separation is much larger than the critical depths of unbinding. According to eq.~(\ref{unbinding}), the critical depth of unbinding from well 1 is $U_{c,1} = b/z_1z_\text{we}$ for $U_2^\text{ef}=0$, and the critical depth of well 2 is $U_{c,2}= b/z_3 z_\text{we}$ for $U_1^\text{ef}=0$. The critical depth $U_c^\text{ef}= c /z_\text{ba}z_\text{we}$ for lateral phase separation is much larger than both $U_{c,1}$ and $U_{c,2}$ since the ratio of the prefactors in the critical scaling laws (\ref{unbinding}) and (\ref{Uc_ef}) is $b/c\simeq 0.11$. For $U_1^\text{ef}=U_2^\text{ef}=U_c^\text{ef}$, the membranes therefore are clearly bound in both wells, which justifies that we have neglected the hard-wall repulsion in deriving eq.~(\ref{Uc_ef}). The resulting phase diagram is shown in Fig.~\ref{figure_phasediagram}. 

In summary, we have characterized the phase behavior of membranes adhering via long and short receptor/ligand bonds. We have mapped the partition function of such membranes on the corresponding partition function of homogeneous membranes interacting via an effective double-well potential, and have determined the critical points of unbinding and lateral phase separation as a function of the characteristic potential parameters. These parameters are the effective depths of the two wells, which depend on the binding energies and concentrations of the receptors and ligands, and the characteristic length scales of the potential, which depend on the dimensions of the receptor/ligand bonds.

\end{document}